\documentstyle[prl,preprint,aps]{revtex}
\draft
\tightenlines
\newcommand{\beq}{\begin{equation}}
\newcommand{\eeq}{\end{equation}}

\begin{document}
\title{Density functional theory versus\\
the Hartree Fock method: comparative assessment}
\author{M.Ya. Amusia$^{a,b}$,
A.Z. Msezane$^{c}$, and V.R. Shaginyan$^{c,d}$
\footnote{E--mail: vrshag@thd.pnpi.spb.ru}}
\address{$^{a\,}$The Racah Institute of Physics,
the Hebrew University, Jerusalem 91904, Israel;\\
$^{b\,}$A.F. Ioffe Physical-Technical Institute, 194021 St.
Petersburg, Russia;\\
$^{c}$CTSPS, Clark Atlanta University, Atlanta,
Georgia 30314, USA;\\
$^{d\,}$Petersburg Nuclear Physics Institute,
Gatchina, 188350, Russia}
\maketitle

\begin{abstract}
We compare two different approaches to investigations of
many-electron systems. The first is the Hartree-Fock (HF) method
and the second is the Density Functional Theory (DFT). Overview of
the main features and peculiar properties of the HF method are
presented. A way to realize the HF method within the Kohn-Sham
(KS) approach of the DFT is discussed. We show that this is
impossible without including a specific correlation energy, which
is defined by the difference between the sum of the kinetic and
exchange energies of a system considered within KS and HF,
respectively. It is the nonlocal exchange potential entering the
HF equations that generates this correlation energy. We show that
the total correlation energy of a finite electron system, which
has to include this correlation energy, cannot be obtained from
considerations of uniform electron systems.  The single-particle
excitation spectrum of many-electron systems is related to the
eigenvalues of the corresponding KS equations. We demonstrate that
this spectrum does not  coincide in general with the eigenvalues
of KS or HF equations.
\end{abstract}


\section{Introduction}

The Hartree-Fock method (HF) originate from the pioneer works of
Hartree \cite{har} and Fock \cite{foc}. Since then an impressive
progress has been achieved by using HF as a basis when calculating
the ground state energy and constructing the Random Phase
Approximation (RPA) and the Random Phase Approximation with
Exchange (RPAE) to investigate the dynamic properties of
multi-electron objects, such as atoms \cite{amu}, molecules,
clusters and fullerenes. Unfortunately, HF method has a number of
difficulties in its application. For instance, one needs a whole
set of single-particle wave functions to calculate the
single-electron nonlocal potential. It is also necessary to follow
a rather intricate procedure to include the correlation
corrections that are beyond the HF framework. As a result, in the
case of complex many-electron systems, starting with molecules,
that include at least several atoms, the calculations become too
complicated.

The Density Functional Theory (DFT) that was invented in the early
1960s has, generally speaking, overcome these difficulties. DFT
has reduced the problem of calculating the ground state
characteristics of a many-electron system in a local external
field to the solution of the Hartree-like one-electron equations
\cite{hwk,wks,parr}. At first DFT was limited to consideration of
only the ground states of many-electron systems, leaving aside
their dynamic properties, which are closely related to the systems
behavior in the time-dependent external fields. This limitation of
DFT was surmounted by Runge and Gross \cite{rg} by transforming
DFT into the so-called Time-Dependent Density Functional Theory
(TDDFT), which in fact is a quite natural generalization of DFT.

Both DFT and TDDFT are based on the one-to-one correspondence
between the particle densities of the considered systems and external
potentials acting upon these particles. For  definiteness, we assume
that the time dependent part of the considered electronic system's
density $\rho({\bf r},t)$, which is created under the action of an
external time-dependent field $\lambda v_{ext}({\bf r},t)$, is
developing from the system's ground state. Consequently,
one can conclude that the wave function and each observable,
describing the system, can be presented as a functional of the
density. Examples are the functionals of the ground state energy
$E[\rho]$ and of the action $A[\rho]$ \cite{wks,parr,rg}.

DFT and TDDFT establish an exact correspondence between a system
of many interacting particles and a fictitious non-interacting
Kohn-Sham (KS) system \cite {wks,parr,rg,gdp}. The main advantage
of this "mapping" is that KS system is described by a set of
Hartree-type single particle equations, which is not too difficult
to solve. As a result, after solving these equations in the case
of an external time-independent potential, one can obtain and
predict, at least in principle, the atomic, molecular, cluster and
solid bodies binding energies, phonon spectra, activation barriers
etc. (see e.g. \cite {parr,gdp}). The same is true in the case of
a time-dependent potential when the solution yields the
single-particle and collective excitation spectra (see e.g.
\cite{gdp,s1,as1,ksk}).

Unfortunately, the one-to-one correspondence proved in
\cite{hwk,rg} establishes the existence of the functionals only in
principle, leaving aside a very important question on how one can
construct them in reality.  This is why the success of DFT and
TDDFT strongly depends upon the availability of good
approximations for the functionals. A systematic way to construct
the required functionals and the corresponding response functions
and effective interaction was suggested in a number of papers
\cite{s1,as1,ksk}.

Since the HF and the Kohn-Sham theory play an important role in
the many-body physics, it deserves comprehensive examinations in
order to reveal their shortcomings and advantages as compared to
one another when applied to studies of the many-electron systems.
Recently, a key element of DFT, the action functional $A[\rho]$,
was suspected to be ill-defined because of a number of
contradictions in it. It appeared that the contradiction, which
came from the analysis of the causality and symmetry properties of
the linear response functions in TDDFT, plays the central role in
creating other contradictions \cite{gdp,bg,l}. Fortunately, these
contradictions were successfully resolved recently
\cite{ksk,asl,camp,hb}.

Exhaustive calculations of the ground state energies of different
atoms within the HF approach and that based on the density
functional corresponding to the HF approximation (DHF) have shown
that there exists an inequality, \begin{equation} E_{DHF}\geq
E_{OEP}>E_{HF},
\end{equation}
where $E_{DHF}$ is the ground state energy calculated in DHF,
using a KS model system with the exact same density as that of
some given HF ground states \cite{goer,nes,ns1}. Here $E_{OEP}$
denotes the energy from the optimized potential method (OPM)
\cite{goer,tsh}, and $E_{HF}$ denotes the HF ground state energy
(see e.g. \cite{goer,nes,ns1}). One can assume that inequality
$E_{OEP}>E_{HF}$ results from the fact that $E_{OEP}$ is
calculated with the restrictions associated with local potentials
while calculations of $E_{HF}$ are in this sense unrestricted (see
e.g. \cite{sahni}). But such considerations could justify only the
relation $E_{OEP}\geq E_{HF}$. For instance, if one applies this
consideration to the functional of the KS kinetic energy
$T_k[\rho]$ one would expect that $T_k[\rho]>T_k^{HF}[\rho]$.
Quite surprisingly, one has instead $T_k^{HF}[\rho]\geq T_k[\rho]$
( see e.g. \cite{goer}). Here $T_k^{HF}[\rho]$ is the HF
functional of the kinetic energy. Then, one could expect, however,
that instead the following relation holds (see, e.g. \cite {nes})
\begin{equation}
E_{DHF}=E_{OEP}=E_{HF}, \end{equation} since a universal
Hohenberg-Kohn (HK) functional is defined for HF ground states
because of the constrained-search derivation \cite{levy}. Note,
that this procedure also does not define explicitly the HF
functional $F^{HF}[\rho]$ whose minimization leads to the HF
ground state. The obvious contradiction between inequality (1) and
equality (2) casts doubts on the reliability of the KS scheme and
probably points to the possible contradictions in DFT based on the
suggestion that an exact local exchange potential does not exist
for ground states of typical atoms (see \cite{nes,ns1} and
references therein). To solve the problem, we have to clarify the
relationships between the HF exchange and kinetic functionals and
the corresponding exchange and kinetic functionals of the DFT.

Without resorting to a numerical analysis, by clarifying the
relationship between the nonlocal exchange HF potential and the
local exchange KS potential, we show that
inequality (1) and equality (2)
have to be replaced by the inequality
\beq E_{OEP}\geq E_{DHF}=E_{HF}. \eeq

We also consider the important problem of how to relate the single
particle spectrum of a many electron system to the corresponding
KS eigenvalues. This consideration enables us to clarify the
relationship between the single particle spectra obtained in HF
calculations and in the KS calculations.

The structure of the paper is the following. In Section II, for
the reader's convenience we review the main features of the HF
method. We list only the main peculiarities of the HF method,
which are directly determined by the nonlocal HF potential. One
cannot expect to reproduce these within the Kohn-Sham theory of
DFT. The explanation of Eqs. (1), (2), and (3) presented in
Section III consists in demonstrating that the contradictions come
from omitting the correlation energy, which is defined by the
difference between the sum of the kinetic and exchange energies of
a system considered within DFT and HF approaches, respectively.
Consequently, this difference, which has to be treated as a
correlation energy $F^{HF}_c[\rho]$, gives rise to a specific
correlation energy, because the nonlocal HF potential ensures the
deeper minimum value of the ground state energy than the local
potential of OPM. We also clarify the relationships between HF
exchange and kinetic functionals and the corresponding exchange
and kinetic functionals of the DFT. Our consideration also shows
that this correlation energy $F^{HF}_c[\rho]$, being the specific
feature of a finite system, cannot be represented by the local
density approximation based on calculations of a homogeneous
electron system. Therefore, the correlation energy of a finite
electron system, which has to include $F^{HF}_c[\rho]$, cannot be
obtained from considerations of uniform electron systems, although
this way to construct $E_c[\rho]$ is generally used \cite{parr}.
In Section IV, we consider the single-particle excitation spectra
of many-electron systems and relate them to the eigenvalues of the
corresponding KS equations. We show that the single-particle
spectra do not coincide with the eigenvalues of the KS or HF
equations. A possible inconsistency in the Kohn-Sham theory based
on calculations of the single-particle spectra \cite{nes,ns1} is
also clarified in this Section. Finally, Section V contains our
concluding remarks.

\section{The Hartree-Fock method}

Hartree's equations, as Fock's generalization of them, were
derived for any atom with two and more electrons. In fact, they
are applicable to any many-electron system in a static external
potential $v({\bf r})$. Assuming that the total electron wave
function $\Psi ({\bf r}_{1},...{\bf r}_{N})$ can be approximated
as a product of one-particle functions $\phi_{i}^{H}({\bf r})$,
$\Psi^{H}({\bf r}_{1},...{\bf
r}_{N})=\prod\nolimits_{i=1}^{N}\phi_{i}^{H}({\bf r}_{i}),$
Hartree suggested \cite{har} the following equation for
$\phi_{i}^{H}({\bf r})$\footnote{ Atomic system of units is used
in this paper: $e=m=\hbar =1$, where $e$ and $ m$ are electron
charge and mass, respectively.}:
\begin{equation}
\left[ -\frac{\Delta }{2}+v({\bf r})+\sum\limits_{j=1,j\neq
i}^{N}\int\frac{|\phi_{j}^{H}({\bf r}^{\prime
})|^{2}}{|{\bf r}-{\bf r}^{\prime }|}d{\bf r}^{\prime }\right]
\phi_{i}^{H}({\bf
r})=E_{i}^{H}\phi_{i}^{H}({\bf r}).  \end{equation}
Here $N$ is the total number of electrons. For an atom one has
$v({\bf r} )=-Z/r$, where $Z$ is the nuclear charge. An electron in
an $i$-state is under the action of the so-called self-consistent field,
which is determined by all electrons but the $i$th one. Excluding the
term $j=i$  from the sum in (4) eliminates the action of the
$i$-electron upon itself. In Hartree approximation the system's
electron density $\rho^{H}({\bf r})$ is determined by the relation
$$
\rho^{H}({\bf r})=\sum\limits_{j=1} n_i |\phi_{j}^{H}({\bf
r})|^{2},$$
where $n_{i}$ are the occupation numbers.

Hartree's approach has two obvious shortcomings: the total
wave-function violates the exclusion principle and, since the
self-consistent potential acting upon the $i$th electron depends
upon $i$, the functions $\phi_{i}^{H}({\bf r})$ are
non-orthogonal. These shortcomings were eliminated in \cite{foc}
by constructing the total wave function as an {\it
anti-symmetrized} $(A)$ product, $$ \Psi^{HF}({\bf r}_{1},...{\bf
r}_{N})=A\prod\limits_{j=1}^{N}\phi _{i}^{HF}( {\bf r}_{i}),$$
which automatically satisfies the exclusion principle. This
wave-function's choice leads to the following ground state energy
$E_{HF}$
\begin{eqnarray}
E_{HF} & = & -\frac{1}{2}\sum\limits_{j=1} n_j\int\phi_j^{\ast}
({\bf r})\nabla^{2}\phi_j({\bf r})d{\bf r}
+\frac{1}{2}\sum\limits_{j=1,i=1} n_j n_i
\int\frac{\phi_j^{\ast}({\bf r}_{1})
\phi_i^{\ast}({\bf r}_{2})\phi_j({\bf r}_{1})
\phi_i({\bf r}_{2})}{|{\bf r}_{1}-{\bf r}_{2}|}d{\bf r}_{1}d{\bf
r}_{2}\\
& -& \frac{1}{2}\sum\limits_{j=1,i=1} n_j n_i
\int\frac{\phi_j^{\ast}({\bf r}_{1})
\phi_i^{\ast}({\bf r}_{2})\phi_i({\bf r}_{1})
\phi_j({\bf r}_{2})}{|{\bf r}_{1}-{\bf r}_{2}|}d{\bf r}_{1}d{\bf
r}_{2}
+\sum\limits_{j=1} n_j\int\phi_j^{\ast}({\bf r})\phi_j({\bf
r})v({\bf r})d{\bf r},\nonumber  \end{eqnarray}
and HF system of
equations:  \begin{eqnarray} \hat{H}^{HF}\phi_{i}^{HF}({\bf r}) &
\equiv & \left[-\frac{\Delta }{2}+v({\bf r})\right]
\phi_{i}^{HF}({\bf r})\\
& + & \sum\limits_{j=1}^{N}\int
\frac{d{\bf r}^{\prime}}
{|{\bf r}-{\bf r}^{\prime }|}
\left[|\phi_{j}^{HF}({\bf r}^{\prime })|^{2}
\phi_{i}^{HF}({\bf r})-\phi_{j}^{HF\ast }({\bf r}^{\prime })
\phi_{i}^{HF}({\bf r}^{\prime
})\phi_{j}^{HF}({\bf r})\right]=
E_{i}^{HF}\phi_{i}^{HF}({\bf r}) \nonumber.
\end{eqnarray}
These equations differ from an ordinary Schr\"{o}dinger equation in
two essential aspects: they are non-linear in $\phi_{i}^{HF}({\bf
r})$ and the second term under the sum that represents the so-called
Fock's potential is non-local.

The non-linearity of Hartree and HF equations leads to
non-uniqueness of their solutions (see \cite{ano} and references
therein): in some cases, a given energy value can be represented
by more than one set of solutions which deliver this value with
high numerical accuracy. Furthermore, the nonlocal HF potential
leads to a number of very specific features of the ground state
wave function, which cannot be reproduced by the KS equations, see
Sec. III.

1. The asymptotic behavior in $r$ of $\phi_{i}^{HF}({\bf r})$, contrary to
the case of an ordinary one-particle Schr\"{o}dinger and also Hartree
equations, is not determined by $E_{i}^{HF}$ and does not behave as
\begin{equation}
\phi_{i}({\bf r})|_{r\rightarrow\infty }\sim\exp
(-\sqrt{2|E_{i}^{HF}|}r).  \end{equation}
On the contrary, it was shown in \cite{hms} that at $r\rightarrow
\infty $ the function $\phi_{i}^{HF}({\bf r})$ is not determined by
$E_{i}^{HF}$ but is of the form \begin{equation}\phi_{i}^{HF}({\bf
r})|_{r\rightarrow\infty }\sim\sum\nolimits_{l}C_{l}[\exp
(-\sqrt{2|E_{F}^{HF}|}r)]/r^{l+1}, \end{equation} where $E_{F}^{HF}$
is the energy of the so-called Fermi-level (with wave function $\phi
_{F}^{HF}({\bf r})$), which is the smallest binding energy among all
$E_{i}$ in the considered system, and \begin{equation}
C_{l}=\sum\limits_{m=-l}^{+l}\int
\phi_{F}^{HF\ast }({\bf r})r^{l}
{\bf Y}_{lm}({\bf r}/r)\phi_{i}^{HF}({\bf r})d{\bf r}.
\end{equation}
Here ${\bf Y}_{lm}({\bf r}/r)$ is the $l^{th}$ order spherical
polynomial. The alteration of the asymptotic behavior dramatically
affects, for example, the probability of penetration of atomic
electrons via a potential barrier created when a strong electric
static field is applied to this atom \cite{amus}.

2. The number of zeroes in the radial part of the function $\phi
_{i}^{HF}( {\bf r})$ is not equal, according to \cite{hfze} and
contrary to the case of an ordinary one-particle Schr\"{o}dinger and
also Hartree equations, to the so-called radial quantum number
$n_{r}=n-l-1$, where $n$ is the principal and $l$ is the angular
momentum quantum numbers. The extra zeroes are located at big
distances and their presence can be explained qualitatively in the
following way. At large distances $\phi_{i}^{HF}({\bf r})$ is
determined mainly by the admixture of $\phi_{F}^{HF}({\bf r})$ to
$\phi_{i}^{HF}({\bf r})$. The latter function adds its remote zeroes
to that of $ \phi_{i}^{H}({\bf r})$.

3. In the presence of the HF potential the electron velocity
operator $\hat{{\bf v}}$, which in quantum mechanics is determined
by the relation $\hat{{\bf v}}=i[\hat{H}^{HF},{\bf r]}$, is not
equal to $\hat{{\bf p}}$ (note, that $m=1$), contrary to the case
of a local potential. This drastically affects, for example, the
photoionization cross section calculations: the amplitude of this
process becomes different in the so-called {\it length} and {\it
velocity} forms and the {\it golden sum rule }is violated
\cite{amu}. It is of interest to note that this violation is very
serious and can be eliminated only when the corresponding
calculations are performed within RPAE framework.

4. One-electron Green's function $G_{E}^{HF}({\bf r}_{1},{\bf
r}_{2})$ in HF has only one expression via $\phi_{i}^{HF}({\bf
r})$, namely \begin{equation} G_{E}^{HF}({\bf r}_{1},{\bf
r}_{2})=\sum\limits_{i}\frac{\phi_{i}^{HF\ast }({\bf
r}_{1})\phi_{i}^{HF}({\bf r}_{2})}{E_{i}-E+i\delta },
\end{equation}
while in case of the spherical symmetry  there exist two
equivalent expressions for local potentials
\begin{equation}
G_{E}(r_{1},r_{2}) =\sum\limits_{i}\frac{\phi _{i}^{\ast
}(r_{1})\phi_{i}(r_{2})} {E_{i}-E+i\delta } =\phi_{E}^{\ast
}(r_{<})\chi_{E}(r_{>}),
\end{equation}
where $r_{>(<)}$ is the larger (smaller) value of
$r_{1,2}$ and $\chi_{E}(r_{>})$ is the solution of the same
Schr\"{o}dinger (and also Hartree) equation as
$\phi_{E}^{\ast}(r_{<})$, but irregular at $r=0$. Note, that
$\phi_{E}(r)$ is regular at $r\rightarrow 0$.

5. Another specific feature of the HF potential is the behavior as
a function of the incoming electron energy $E$ of elastic
scattering phase shifts $\delta_{l}(E)$ of a partial wave with
angular momentum $l$. As is well known, if the wave function's
phase of electron scattering upon a local potential $v({\bf r})$
is normalized in such a way that
$\delta_{l}(E)|_{E\rightarrow\infty }\rightarrow 0$, the relation
$\delta_{l}(0)=q_{l}\pi $ holds, where $q_{l}$ is the number of
the electron's {\it vacant} bound states with angular momentum $l$
in this potential (see e.g.\cite{lali}). The situation for the HF
potential is qualitatively different.  With the same normalization
$\delta_{l}^{HF}(E)|_{E\rightarrow\infty }\rightarrow 0$ another
relation holds: $\delta_{l}^{HF}(0)=(q_{l}+t_{l})\pi$, where
$t_{l}$ is the number of {\it occupied} bound states with angular
momentum $l$ in the HF potential \cite{amu,ano}.

\section{The KS local exchange potential and the Hartree-Fock
nonlocal one}

Consider a many-body system composed of $N$ interacting electrons
and moving in an external potential $v({\bf r})$, with
Hamiltonians $\hat{H}$ and $\hat{H_{v}}$ that are of the form
\begin{equation} \hat{H}=\hat{T}+\hat{U}, \end{equation} and
\begin{equation} \hat{H_{v}}=\hat{T}+\hat{U}+\hat{V}.
\end{equation}
Here $\hat{T}$, $\hat{U}$, and $\hat{V}$ are given in terms of the
field operators $\hat{\psi }({\bf r})$, and $\hat{T}$ is the kinetic
energy operator \begin{equation}
\hat{T}=-\frac{1}{2}\int\hat{\psi}^{\ast}
({\bf r})\nabla^{2}\hat{\psi}({\bf r})d{\bf r},
\end{equation}
$\hat{U}$ is the interelectron interaction energy operator
\begin{equation}
\hat{U}=\frac{1}{2}\int\frac{\hat{\psi}^{\ast}({\bf r}_{1})\hat{
\psi}^{\ast}({\bf r}_{2})\hat{\psi}({\bf r}_{2})\hat{\psi}({\bf
r}_{1})}{|{\bf r}_{1}-{\bf r}_{2}|}d{\bf r}_{1}d{\bf r}_{2},
\end{equation}
and $\hat{V}$ is the potential energy operator in the field
$v({\bf r})$ \begin{equation} \hat{V}=\int\hat{\psi}^{\ast}({\bf
r})\hat{ \psi}({\bf r})v({\bf r})d{\bf r}.  \end{equation}
Hohenberg and Kohn proved that there exists a one-to-one
correspondence between $v({\bf r})$ and the single-electron
density $\rho({\bf r})$, that is $v({\bf r})$ is a unique
functional of the density, apart from an additive constant
\cite{hwk}. Because $v({\bf r})$ fixes the
Hamiltonian Eq. (13), the ground
state wave function $\Psi_{0}$ is a functional of the density, and
one can define the energy functional \cite{hwk}
\begin{equation} E[\rho]=F[\rho]+\int v({\bf r})\rho({\bf r})d{\bf
r},
\end{equation} with $F[\rho]$ being a universal functional
independent of the number of particles $N$ and the external
potential $v({\bf r})$. Note that the functional $F[\rho]$ can
also be obtained on the basis of the constrained-search derivation
of Levy, when it is defined by the minimum over the set of all
normalized $N$-electron wave functions $\Psi_{s}$ that determine
the density $\rho({\bf r})$ \cite{levy}
\begin{equation}
F[\rho]=\min\left(\Psi_{s}|\hat{T}+\hat{U}|\Psi_{s}\right).
\end{equation}
Provided the density is exact, $\rho({\bf r})=\rho_{0}({\bf
r})$, and normalized by the number of particles $N$,
$N=\int\rho({\bf r})d{\bf r}$, the functional $E[\rho]$ assumes its
minimum value that is equal to the ground state energy $E_{0}$
\begin{equation}
E_{0}=F[\rho_{0}]+\int v({\bf r})\rho_{0}({\bf r})d{\bf r}.
\end{equation}
Here the ground state density $\rho_{0}({\bf r})$ is determined by
the equation \begin{equation} \frac{\delta
F[\rho]}{\delta\rho({\bf r})}+v({\bf r})=\mu, \end{equation} and
$\mu $ is the so-called chemical potential. The number $\mu $
ensures the conservation of the number of particles $N$. It is
clear that if $\hat{U}=0$, [see Eq. (13)], the functional
$F[\rho]$ reduces to the functional $T_{k}[\rho]$ of the kinetic
energy of noninteracting particles.

Equation (20) can be replaced by a set of Hartree-like
single-particle equations [see Eq. (4)],
which permits the calculation of the
density $\rho({\bf r})$ \cite{wks} \begin{equation} \left(
-\frac{\bigtriangleup }{2}+V_{L}({\bf r})\right)\phi_{i}({\bf r}
)=\varepsilon_{i}\phi_{i}({\bf r}),
\end{equation}
with the potential $V_{L}=V_{H}+V_{xc}+v$. The density $\rho({\bf r})$
is given by \begin{equation} \rho({\bf
r})=\sum_{i}n_{i}|\phi_{i}({\bf r})|^{2}.  \end{equation}
Here $V_{H}$ is the Hartree potential [see Eq. (4)],
\begin{equation}
V_{H}({\bf r})=\frac{\delta }{\delta\rho({\bf r})}F_{H}=\int
\frac{\rho( {\bf r}_{1})}{|{\bf r}-{\bf r}_{1}|}d{\bf r}_{1},
\end{equation}
where $F_{H}$ is the Hartree functional
\begin{equation}
F_{H}=\frac{1}{2}\int\frac{\rho({\bf r}_{1})\rho({\bf
r}_{2})}{|{\bf r}- {\bf r}_{1}|}d{\bf r}_{1}d{\bf r}_{2},
\end{equation}
and the exchange-correlation potential,
$V_{xc}({\bf r})=\delta F_{xc}/\delta\rho({\bf r})$,
is obtained from the exchange-correlation functional,
\begin{equation}
F_{xc}[\rho]=F[\rho]-T_{k}[\rho]-F_{H}[\rho].  \end{equation} In
the same way, it is possible to define separately $F_{x}$ and the
correlation functional $F_{c}$ using the equation, which can be
considered as a definition of $F_{c}$ \cite{wks} \begin{equation}
F_{c}[\rho]=F_{xc}[\rho ]-F_{x}[\rho].  \end{equation} Here
$F_{x}[\rho]$ is the exchange functional \cite{as1,s2}
\begin{eqnarray}
F_{x}[\rho] & = & -\frac{1}{2}\int
\frac{\chi_{0}({\bf r}_{1},{\bf r}_{2},iw)+2\pi
\rho({\bf r}_{1})\delta(w)\delta({\bf r}_{1}-{\bf r}_{2})}{|{\bf
r}_{1}-{\bf r}_{2}|}\frac{dw}{2\pi}d{\bf r}_{1}d{\bf r}_{2}\\
& = & -\sum_{k,i}n_{k}n_{i}\int\left[\frac{\phi_{i}^{\ast }({\bf
r}_{1})\phi_{i}({\bf r}_{2})\phi_{k}^{\ast }({\bf r}_{2})\phi
_{k}({\bf r}_{1})}{|{\bf r}_{1}-{\bf r}_{2}|}\right] d{\bf
r}_{1}d{\bf r}_{2} \nonumber,
\end{eqnarray} and $\chi_{0}$ is the linear response
function, which is of the form \begin{equation} \chi_{0}({\bf
r}_{1},{\bf r}_{2},\omega )=\sum_{i,k}n_{i}(1-n_{k})\phi_{i}^{\ast
}({\bf r}_{1})\phi_{i}({\bf r}_{2})\phi_{k}^{\ast }({\bf r}
_{2})\phi_{k}({\bf r}_{1})\left[\frac{1}{\omega -\omega_{ik}+i\eta }-
\frac{1}{\omega +\omega_{ik}-i\eta }\right],
\end{equation}
with $\omega_{ik}$ defined as $\omega_{ik}=\varepsilon
_{k}-\varepsilon_{i}$, $\varepsilon_{k}$ and functions $\phi
_{k}({\bf r}_{1})$ being respectively the one-particle energies and
wave functions of Eq. (21); $\eta\rightarrow 0$. Having the
functional $F_{x}$ given by Eq. (27), we can construct the HF-like
scheme based on Eq. (21) with the local KS exchange potential
\cite{as1,s2} \begin{equation} V_{x}({\bf r})=\frac{\delta }{\delta
\rho({\bf r})}F_{x}[\rho].  \end{equation}
The functional of density $F_{x}^{KS}$ corresponding to the HF
approximation can be defined as follows
\begin{equation}
F_{x}^{KS}[\rho]=T_{k}[\rho]+F_{H}[\rho]+F_{x}[\rho].
\end{equation}
In fact, this functional is determined by the value of
$F_{x}^{KS}[\rho]=(\Phi^{KS}|\hat{H}|\Phi^{KS})$, if $\Phi^{KS}$
is a single $N$-electron Slater determinant which delivers the
lowest energy expectation value of $\hat{H}$ given by Eq. (12) and
yielding the density $\rho({\bf r})$. This determinant is composed
of the single-particle wave functions, which are solutions of a
single-particle equation like Eq. (21) provided that $V_{L}( {\bf
r})$ is a local potential. We denote this determinant by $\Psi
^{KS}[\rho]$. On the other hand, the same wave function in the
frame of the optimized effective potential \cite{tsh} is
determined by the local exchange potential $V_{OPM}$. Because of
the one-to-one correspondence between this wave function and the
local single-particle potential \cite{hwk} we can conclude that
$V_{OPM}$ coincides with $V_{x}$ given by Eq. (29). This result is
in accord with the direct comparison of $ V_{x}$ with $V_{OPM}$
that confirms their equality, $V_{x}({\bf r})=V_{OPM}({\bf r})$
\cite{as1,s2}. Therefore, the ground state energy $E_{OEP}$ of a
many-electron system calculated in OPM has to be equal to the
corresponding energy calculated with the functional given by Eq.
(30), $E_{KSX}$, that is $E_{KSX}=E_{OEP}$.

Now we turn to the HF calculations based on the constrained-search
formulation of DFT, which leads to the HF functional
$F^{HF}[\rho]$ \cite{levy}. As it was mentioned in Section II and
in accord with the constrained-search formulation \cite{levy}, the
HF functional is determined by the minimum of the expectation
value $F^{HF}[\rho ]=(\Psi\lbrack\rho]|\hat{H}|\Psi\lbrack\rho])$
when the minimization runs over all single-electron Slater
determinants yielding the density $\rho({\bf r})$.  The
determinant, which yields the minimum value, is denoted as
$\Psi_{HF}[\rho]$. This construction of the functional
$F^{HF}[\rho]$ makes it clear that the ground state energy
$E_{DHF}$ of an electron system calculated with $F^{HF}[\rho]$ is
equal to $E_{HF}$. Note that the determinant $\Psi_{HF}[\rho]$ is
composed of the eigenfunctions of Eq. (6) incorporating the Fock
nonlocal potential. As a result, we can at first define the HF
functional of the kinetic energy
\begin{equation}
T_{k}^{HF}[\rho]=\left(\Psi_{HF}[\rho]|\hat{T}|\Psi_{HF}[\rho
]\right), \end{equation} and then the HF functional of the
exchange energy \begin{equation}
F_{x}^{HF}[\rho]=\left(\Psi_{HF}[\rho
]|\hat{U}|\Psi_{HF}[\rho]\right)-F_H[\rho]. \end{equation}

We will now show that $T_{k}^{HF}[\rho]\neq T_{k}[\rho]$, and our
proof proceeds by {\it reductio ad absurdum}. Note that there is a
one-to-one correspondence between $\rho({\bf r})$, the
single-particle potential and the ground state wave function
\cite{hwk}. Because $T_{k}[\rho]$ is a unique functional of the
density, each value of $T_{k}[\rho]$ fixes the only ground state
function. This function is a single Slater determinant $\Phi^{KS}$
yielding the density $\rho({\bf r})$ normalized to $N$ electrons
and characterized by the single-particle wave functions $\phi_{i}$
and eigenvalues $\varepsilon_{i}$, given by Eq. (21). Assume that
$T_{k}^{HF}[\rho]=T_{k}[\rho].$ Then the HF wave functions $\Psi
^{HF}[\rho]$ and the eigenvalues $E^{HF}_i$ given by Eq. (6) must
be equal to $\Phi ^{KS}[\rho]$ and the eigenvalues $\varepsilon_i$
given by Eq. (21). As it was mentioned in Sec. II, Eq. (8), all of
the HF single-particle functions have the same long-range behavior
which is dominated by the smallest orbital energy \cite{hms},
while, generally speaking, the eigenvalues $E^{HF}_i$ are not
degenerate. On the other hand, eigenfunctions of Eq. (21) can have
such a behavior only if the eigenvalues $\varepsilon_i$ are
degenerate. In order to eliminate the contradiction we have to
recognize that $T_{k}^{HF}[\rho]\neq T_{k}[\rho]$, and some of
densities, which are the ground state densities of single-particle
Schr\"{o}dinger equations with nonlocal potentials are not
noninteracting $v$ representable in the sense that they cannot be
the ground state density of Eq. (21) with some local potential
$v({\bf r})$. Note that examples of densities which are not $v$
representable were given in Ref. \cite{levy1}. Obviously, a
one-to-one correspondence between nonlocal potentials and local
ones does not exist \cite{tlg}. Therefore, if we would even have
the functional $F^{HF}[\rho]$ explicitly we would not have been
able to calculate the HF ground state within  the KS theory
because the HF ground state cannot be reproduced by Eq. (21) with
a local single-particle potential. On the other hand, the HF
ground state can be obtained by solving the Hohenberg-Kohn
equation (20) with $F[\rho]=F^{HF}[\rho]$. These facts do not
imply any contradictions. The Kohn-Sham theory and Eq. (20) lead
to the same result if the density $\rho({\bf r})$ of system under
consideration is $v$ representable. Otherwise one has to use Eq.
(20). A detailed investigation of this important problem will be
published elsewhere.

The functional $T_{k}^{HF}$ given by Eq. (31) is determined in the
domain of the HF densities $\rho_{HF}$ generated by the non-local
Fock potential entering Eq. (6), while $T_{k}$ is determined  in
the domain of the densities $\rho_L$ given by Eq. (21). Note that
both of the densities are normalized to same number of electrons.
Approximating  $\rho_{HF}$ by $\rho_{L}$, one has to obtain
$T_{k}< T^{HF}_{k}$ because of the minimum principle which ensures
that the Slater determinant $\Phi^{KS}$ delivers the minimum value
of the kinetic energy $T_k$. This result is confirmed by numerical
calculations \cite{goer}. On the other hand, due to the minimum
principle $F^{HF}<F^{KS}_x$. Following Ref.\cite{goer}, we can
define the HF exchange-correlation functional $F_{xc}^{FH}[\rho]$
\begin{equation}
F_{xc}^{FH}=F_{x}^{HF}-T_{k}-F_{H} =T_{k}^{HF}+F_{x}^{HF}-T_{k},
\end{equation}
and introduce the HF correlation functional $F_{c}^{FH}$
\begin{equation} F_{c}^{FH}=F_{xc}^{FH}
-F_{x}=T_{k}^{HF}+F_{x}^{HF}-T_{k} -F_{x}.  \end{equation} We note
that Eqs. (33) and (34) cannot be considered as exact definitions
of $F_{xc}^{FH}$ and  $F_{c}^{FH}$ because it is impossible to
approximate exactly $\rho_{HF}$ by $\rho_{L}$. It follows from Eq.
(34) that the correlation energy $F_c^{HF}$ exists. This energy
must be taken into account when constructing constructing
approximations of $F^{HF}$ within the KS framework. Omitting
$F_c^{HF}$, one obtains Eq. (1), while properly constructing
$F^{HF}[\rho]$ based on the constrained-search formulation of DFT
\cite{levy}, one obtains Eq. (3) rather then Eq. (2).

In the case of He atoms the HF potential acting on the occupied
states is a local potential. Therefore, the kinetic energy
functional $T^{HF}_k[\rho]$ coincides with $T_k[\rho]$ leading to
the equality $ F_{x}^{HF}[\rho ]=F_{x}^{KS}[\rho]$, and
$E_{OEP}=E_{DHF}=E_{HF}$ in line with Eqs. (2) and (3). This
observation is confirmed by numerical calculations \cite{nes}. For
all other atoms, the HF potential is nonlocal, and
$F^{HF}_c[\rho]$ contributes to the ground state energy.
Obviously, one has $E_{OEP}\geq E_{DHF}=E_{HF}$ in accordance with
Eq. (3) and the numerical calculations \cite{nes}. In the case of
a homogeneous system, both single-particle wave functions, $\phi
_{i}^{HF}({\bf r})$ given by Eq. (6) and the ones $\phi_{i}({\bf
r})$ given by Eq. (21), are plane waves. As a result, the energy
$E_{KSX}=E_{HF}$ (though the Fock potential, Eq. (6), is a
nonlocal) and the correlation energy $E_{c}=E-E_{KSX}=E-E_{HF}$ is
uniquely defined. This consideration shows that $F^{HF}_c[\rho]$,
being the specific feature of a finite system, cannot be
approximated by the local density approximation based on
calculations of a homogeneous electron system. Thus, $E_c[\rho]$
of a finite electron system, which has to include
$F^{HF}_c[\rho]$, cannot be obtained from considerations of
uniform electron systems, although this way to construct
$E_c[\rho]$ is generally used \cite{parr}. As a result, we are led
to the conclusion that the HF method has unique peculiarities
which can be considered as useful rather then as shortcomings when
investigating many-electron systems. Moreover, it is of great
importance to check in experiments whether the wave function of a
real physical system has the behavior resembling the behavior of
the HF wave function given by Eq. (8) because this point is
closely related to the problem of the $v$-representability. As
mentioned in Sec. II, this HF-like  behavior can be checked
measuring  the probability of penetration of atomic electrons via
a potential barrier \cite{amus}.

\section{ Single particle spectra}

In order to calculate the single-particle energies $\epsilon_{i}$
of a many-electron system we have to consider the ground state
energy as a functional of the density $\rho$ and of the occupation
numbers $n_{i}$. The occupation numbers are widely used in
generalizations of DFT when extending DFT to finite temperatures,
when considering excitations energies and so on, for a short
review see Ref. \cite{levy1,nesbet}. Taking into account $n_i$, we
can use the well-known Landau equation of the Fermi liquid theory
\cite{ll} \begin{equation}\frac{\delta E}{\delta n_{i}}=\epsilon
_{i}.\end{equation} We note that there exist close relations
between the Fermi-liquid theory and DFT \cite{ksk}, while the
Fermi-liquid theory is applicable in principle essentially to all
types of Fermi systems including finite objects such as atoms,
molecules, clusters and fullerenes. It is difficult to provide an
example of a Fermi system, which cannot at all be described in the
frame of this theory (see e.g. \cite{ll,var}).

In the functional $E[\rho,n_{i}]$ both of the parameters, $\rho$
and $ n_{i}$, are independent variables and determine different
features of the system under consideration. The ground state
energy as a functional of the density determines the system's
behavior in an external field, while the ground state energy as a
functional of the occupation numbers characterizes the
single-particle spectrum. On the other hand, the functional
$F[\rho]$ given by Eq. (18) depends only on the density. One can
use some approximations to construct $E[\rho,n_{i}]$. For
instance, writing the density in explicit form given by Eq. (22),
we find that the functional $F[\rho]$ depends upon the occupation
numbers $n_{i}$ \cite{nes,ns1,jk}
\begin{equation}
F[\rho,n_{i}]\equiv F\left[\sum_{i}n_{i}|\phi_{i}({\bf
r})|^{2}\right],  \end{equation} while $E[\rho,n_{i}]$ is given by
Eq. (17). In that case, using Eq. (35) one can show that the
single-particle energies $\epsilon_{i}$ coincide with the
eigenvalues $\varepsilon_i$ of Eq.  (21) \cite{jk}.

Let us show that the procedure based on Eq. (36) may lead,
however, to wrong results. Consider a large uniform system of
fermions, so that it is possible to introduce the density of
states at the Fermi-level. In this case, the single-particle
potential is a constant, the eigenvalues of Eq. (21) are the
eigenvalues of noninteracting fermions with bare mass $M$, and the
density of states is equal to that of noninteracting fermions.
This result is incorrect being related to the fact that the
functional form $E[\rho,n_{i}]$ given by Eq. (36) was artificially
chosen. In contrast, as it is seen from Eq. (5), the HF ground
state energy depends on the occupation numbers quite naturally and
leads to the correct results when calculating the single-particle
spectra \cite{ll,bar,gell}. Inserting Eq. (5) into Eq. (35), we
can calculate the HF single-particle spectrum $\epsilon^{HF}_i$
that coincides with the eigenvalues of Eq. (6). In case of a large
uniform system of electrons, one can check that the density of
states tends to zero \cite{bar,gell}. Upon comparing this result
with the result of the KS scheme, one may suspect an inconsistency
in the KS approach. An inconsistency like this was indeed
suspected in KS approach because $\sum_i n_i \epsilon_i\neq \sum_i
n_i E^{HF}_i$ \cite{nes,ns1}, with $\epsilon_i$ being calculated
in the scheme of \cite{jk}.

In order to show that the inconsistencies mentioned above are
fictitious and to illustrate how to calculate the single-particle
energies $\epsilon_{i}$ within the DFT, we choose the simplest
case when the functional is given by Eq. (30). As we shall see,
the single-particle energies $\epsilon_{i}$  coincide neither with
the eigenvalues $E_{i}^{HF}$ of Eq. (6) nor with $\varepsilon_{i}$
of Eq. (21). To proceed, we use a method developed in
\cite{camp,s2,as3,s3}. The linear response functions $\chi_{0}$,
and the density $\rho({\bf r})$ depend upon the occupation
numbers. Thus, one can consider the ground state energy as a
functional of the density and the occupation numbers
\begin{equation} E[\rho({\bf r}),n_{i}]=T_{k}[\rho ({\bf
r}),n_{i}]+F_{H}[\rho({\bf r} ),n_{i}]+F_{x}[\rho({\bf
r}),n_{i}]+\int V_{ext}({\bf r})\rho({\bf r})d {\bf r}.
\end{equation} Here $T_{k}$ is the functional of the kinetic
energy of noninteracting KS particles. Substituting Eq. (37) into
Eq. (35) and remembering that the single-particle wave functions
and eigenvalues are given by Eq. (21), we see that the single
particle spectrum $\epsilon _{i}$ can be represented by the
following expression: \begin{equation}
\epsilon_{i}=\varepsilon_{i}-<\phi_{i}|V_{x}|\phi_{i}>-\frac{1}{2}\frac{
\delta }{\delta n_{i}}\int\left[\frac{\chi_{0}({\bf r}_{1},{\bf r}
_{2},iw)+2\pi\rho({\bf r}_{1})\delta (w) \delta ({\bf r}_{1}-{\bf
r}_{2})} {|{\bf r}_{1}-{\bf r}_{2}|}\right] \frac{dwd{\bf
r}_{1}d{\bf r}_{2}}{2\pi }.\end{equation} The first and second
terms on the right hand side in Eq. (38) are determined by the
derivative of the functional $T_{k}$ with respect to the
occupation numbers $n_{i}$. To calculate the derivative we
consider an auxiliary system of non-interacting particles in a
field $U({\bf r})$. The ground state energy $E_{0}^{U}$ of this
system is given by the following equation
\begin{equation} E_{0}^{U}=T_{k}+\int
U({\bf r})\rho({\bf r})d{\bf r}.  \end{equation} Varying
$E_{0}^{U}$ with respect to the occupation numbers, one gets the
desirable result
\begin{equation}
\varepsilon_{i}=\frac{\delta E_{0}^{U}}{\delta n_{i}}=\frac{\delta
T_{k}}{\delta n_{i}}+<\phi_{i}|U|\phi_{i}>, \end{equation}
provided $U=V_{H}+V_{x}+v$. The third term on the right hand side
of Eq.  (38) is related to the contribution coming from $F_{x}$
defined by Eq. (27). In the considered simplest case when we take
into account only $T_k$, $F_H$ and $F_x$ functionals, the
infraction enters as a linear factor, with $F_H$ and $F_x$ being
linear dependent on the strength. If we omit the inter-electron
interaction given by Eq. (15) we directly get from Eq. (37)
$\epsilon_i=\varepsilon_i$ as it must be in the case of
noninteracting system of electrons. Note that it is not difficult
to include the correlation energy in the simplest local density
approximation
\begin{equation} F_{c}[\rho,n_{i}]=\int\rho({\bf
r})\varepsilon_{c}(\rho({\bf r}))d{\bf r }. \end{equation} Here
the density $\rho({\bf r})$ is given by Eq. (22) and the
correlation potential is defined as
\begin{equation}
V_{c}({\bf r})=\frac{\delta F_{c}[\rho]}{\delta\rho({\bf r})}.
\end{equation}
Then Eq. (37) takes the form
\begin{equation}
E[\rho({\bf r}),n_{i}]=T_{k}[\rho({\bf r}),n_{i}]+F_{H}[\rho({\bf
r} ),n_{i}]+F_{x}[\rho({\bf r}),n_{i}]+F_{c}[\rho({\bf
r}),n_{i}]+\int V_{ext}({\bf r})\rho({\bf r})d{\bf r}.
\end{equation}
Varying $E[\rho({\bf r}),n_{i}]$ with respect to the occupation
numbers $ n_{i}$ and after some straightforward calculations, we
obtain the rather simple expression for the single particle
spectrum
\begin{equation}
\epsilon_{i}=\varepsilon_{i}-<\phi_{i}|V_{x}|\phi
_{i}>-\sum_{k}n_{k}\int\left[\frac{\phi_{i}^{\ast }({\bf
r}_{1})\phi_{i}({\bf r}_{2})\phi_{k}^{\ast }({\bf r}_{2})\phi
_{k}({\bf r}_{1})}{|{\bf r}_{1}-{\bf r}_{2}|}\right] d{\bf
r}_{1}d{\bf r}_{2}.  \end{equation} Here we employ Eq. (40) and
choose the potential $U$ as $U=V_{H}+V_{x}+V_{c}+v$ to calculate
the derivative $\delta T_{k}/\delta n_{i}$. Approximating the
correlation functional $F_{c}[\rho,n_{i}]$ by Eq. (41) we simplify
the calculations a lot, preserving at the same time the asymptotic
condition, $(V_{x}+V_{c})_{r\rightarrow\infty }\rightarrow -1/r$.
This condition is of crucial importance when calculating the wave
functions and eigenvalues of vacant states in the frame of the KS
approach \cite{s2,as3}. Note, that these functions and eigenvalues
that enter Eq. (44) determine the single particle spectrum
$\epsilon_i$.  This spectrum has to be compared with the
experimental results.

Single particle levels $\epsilon_{i}$ given by Eq. (44) resemble
the ones that are obtained using HF method. If the wave functions
$\phi_{i}$ were solutions of the HF equations, the correlation
potential $ V_{c}$ would be omitted, then the energies
$\epsilon_{i}$ would exactly coincide with the eigenvalues
$\varepsilon_{i}$. But this is not the case, since $\phi_{i}$ and
$\varepsilon_{i}$ are the eigenfunctions and eigenvalues of Eq.
(21). Therefore, the energies $\epsilon_{i}$ coincide neither with
HF eigenvalues given by Eq. (6), nor with the ones of Eq. (21). As
we have seen in Section III, in case of the He atom
$T^{HF}_k[\rho]=T_k[\rho]$ and $F^{HF}_x[\rho]=F^{KS}_x[\rho]$.
Then it follows from Eq. (44) that $\sum_i n_i \epsilon_i=\sum_i
n_i E^{HF}_i=\sum_i n_i \varepsilon_i$ in accordance with
numerical results obtained in \cite{nes}. In case of other atoms
we cannot expect this equality to hold because of the non-locality
of the HF single-particle potential. This result is in agreement
with the numerical calculations \cite{nes,ns1} and has nothing to
do with contradictions between the HF method and DFT.

It is worth noting that if the exchange functional $F_{x}[\rho]$
is treated within the local density approximation, as it is done
for $F_{c}[\rho]$, that is one uses Eq. (36) to construct
$E[\rho,n_i]$, then the second and the third terms on the right
hand side of Eq. (44) cancel each other, leading to
$\epsilon_{i}=\varepsilon_{i}$ in accordance with \cite{jk}.
Therefore it is of crucial importance to keep the proper
representation of the functional $F_{x}[\rho]$ given by Eq.  (27).
We also anticipate that Eq. (44) when applied to calculations of
many-electron systems such as atoms and molecules will produce
reasonable results for the energy gap separating the occupied and
empty states. In case of solids, we expect that the energy  gap at
various high-symmetry points in the Brillouin zone of
semiconductors and dielectrics can also be reproduced.

\section{Concluding remarks}

We have shown that there are no contradictions in DFT related to
the fact that the HF-like calculations of finite electron systems
carried out within the Kohn-Sham theory of DFT lead to the ground
state energy, the absolute value of which is smaller than that
calculated within the HF method. On the other hand, we have
demonstrated that the properly constructed HF functional of the
density produces the ground state energy, which coincides with the
energy obtained within HF calculations. While inequalities (1) and
(2) came from omitting the correlation energy, which is defined by
the difference between the sum of the kinetic and exchange
energies of a system considered within the KS approach and the HF
method, respectively. It is the nonlocal exchange potential
entering the HF equations that generates this correlation energy.
We have shown that this functional $F^{HF}_c$ of the correlation
energy, being the specific feature of a finite system, cannot be
approximated by the local density approximation based on
calculations of a homogeneous electron system. Thus, the
correlation energy of a finite electron system, which has to
include $F^{HF}_c[\rho]$, cannot be obtained from considerations
of uniform electron systems, although this way of constructing
$E_c[\rho]$ is generally used. We can conclude that the HF method
has unique peculiarities which can be useful when investigating
many-electron systems.

We have clarified the relationship between the eigenvalues of the
single-particle KS equations, the HF eigenvalues and the real
single-particle spectrum. Again, there are no contradictions based
on the comparison of the single-particle spectra calculated within
DFT and the HF method. We have presented a simple equation
defining the single-particle spectrum of a many-electron system.
It is worth noting that all the specific features of the HF method
mentioned in Section II are dictated by the HF nonlocal potential
and therefore give no grounds to seek possible contradictions in
the Kohn-Sham theory which deals with local single-particle
potentials and $v$ representable densities. As it was mentioned in
Section II, the HF method has some drawbacks. Among them there is
a well-known consequence of the non-locality of the HF
single-particle exchange potential, which leads to the fact that
the HF single-particle potential acting on the unoccupied states
falls of exponentially. As a result, the HF potential can support
very few unoccupied states. Such a poor situation with the number
of states develops difficulties in treating the excited states. By
contrast, the KS theory does not suffer from the drawbacks of the
HF method.

\section{Acknowledgments}

The visit of VRS to Clark Atlanta University has been supported by
NSF through a grant to CTSPS. MYaA is grateful to the S.A.
Shonbrunn Research Fund for support of his research. AZM is
supported by US DOE, Division of Chemical Sciences, Office of
Basic Energy Sciences, Office of Energy Research. This work was
supported in part by the Russian Foundation for Basic Research,
project no. 01-02-17189.

\end{document}